\documentclass[english,aps,preprint]{revtex4}
\usepackage[T1]{fontenc}
\usepackage[latin9]{inputenc}
\usepackage{float}
\usepackage{textcomp}
\usepackage{amsmath}
\usepackage{graphicx}

\makeatletter

\providecommand{\tabularnewline}{\\}

\@ifundefined{textcolor}{}
{%
 \definecolor{BLACK}{gray}{0}
 \definecolor{WHITE}{gray}{1}
 \definecolor{RED}{rgb}{1,0,0}
 \definecolor{GREEN}{rgb}{0,1,0}
 \definecolor{BLUE}{rgb}{0,0,1}
 \definecolor{CYAN}{cmyk}{1,0,0,0}
 \definecolor{MAGENTA}{cmyk}{0,1,0,0}
 \definecolor{YELLOW}{cmyk}{0,0,1,0}
}

\makeatother

\usepackage{babel}
\begin{document}

\title{FCC based $ep$ and $\mu p$ colliders}

\author{Y. C. Acar}

\email{ycacar@etu.edu.tr}

\selectlanguage{english}%

\affiliation{TOBB University of Economics and Technology, Ankara, Turkey }

\author{U. Kaya}

\email{ukaya@etu.edu.tr}

\selectlanguage{english}%

\affiliation{TOBB University of Economics and Technology, Ankara, Turkey }

\affiliation{Department of Physics, Faculty of Sciences, Ankara University, Ankara,
Turkey}

\author{B. B. Oner}

\email{b.oner@etu.edu.tr}

\selectlanguage{english}%

\affiliation{TOBB University of Economics and Technology, Ankara, Turkey }

\author{S. Sultansoy}

\email{ssultansoy@etu.edu.tr}

\selectlanguage{english}%

\affiliation{TOBB University of Economics and Technology, Ankara, Turkey }

\affiliation{ANAS Institute of Physics, Baku, Azerbaijan}
\begin{abstract}
Construction of future electron-positron colliders (or dedicated electron
linac) and muon colliders close to Future Circular Collider will give
opportunity to utilize highest energy proton and nucleus beams for
lepton-hadron and photon-hadron collisions. In this paper we estimate
main parameters of the FCC based $ep$ and $\mu p$ colliders.
\end{abstract}
\maketitle

\section{introduction}

During last decades colliders provide most of our knowledge on fundamental
constituents of matter and their interactions. Particle colliders
can be classified concerning center-of-mass energy, colliding beams
and collider types:
\begin{itemize}
\item Center-of-mass energy: energy frontiers and particle factories,
\item Colliding beams: hadron, lepton and lepton-hadron colliders,
\item Collider types: ring-ring, linac-linac and linac-ring.
\end{itemize}
The ring-ring colliders are most advanced from technology viewpoint
and are widely used around the world. As for the linac-linac colliders,
essential experience is handled due to SLC operation and ILC/CLIC
related studies. The linac-ring colliders are less familiar (for history
of linac-ring type proposals see \cite{mycitation}). 

In Table I we present correlations between colliding beams and collider
types for energy frontier colliders. Concerning the center-of-mass
energy: hadron colliders provide highest values (for this reason they
are considered as \textquotedbl{}discovery\textquotedbl{} machines),
while lepton colliders have an order smaller $E_{CM}$ (for this reason
they are considered as \textquotedbl{}precision\textquotedbl{} machines),
and lepton-hadron colliders provide intermediate $E_{CM}$. It should
be mentioned that differences in center-of-mass energies become fewer
at partonic level. From the BSM search point of view, lepton-hadron
colliders are comparable with hadron colliders for a lot of new phenomena
(for \textquotedbl{}finger estimations\textquotedbl{} see \cite{key-2,key-3}).

\begin{table}[H]
\caption{Energy frontier colliders: colliding beams vs collider types}

\begin{centering}
\begin{tabular}{|c|c|c|c|}
\hline 
Colliders & Ring-Ring & Linac-Linac & Linac-Ring\tabularnewline
\hline 
\hline 
Hadron & + &  & \tabularnewline
\hline 
Lepton ($e^{-}$$e^{+}$) &  & + & \tabularnewline
\hline 
Lepton ($\mu^{-}$$\mu^{+}$) & + &  & \tabularnewline
\hline 
Lepton-hadron ($eh$) &  &  & +\tabularnewline
\hline 
Lepton-hadron ($\mu h$) & + &  & \tabularnewline
\hline 
Photon-hadron &  &  & +\tabularnewline
\hline 
\end{tabular}
\par\end{centering}

\centering{}
\end{table}

\[
\,
\]

Below we list past and future energy frontier colliders for three
time periods:
\begin{itemize}
\item Before the LHC (<2010): Tevatron , SLC/LEP ($e^{-}$$e^{+}$) and
HERA ($ep$),
\item LHC era (2010-2030): LHC ($pp$, $AA$), ILC ($e^{-}$$e^{+}$), low
energy MC ($\mu^{-}$$\mu^{+}$), LHeC ($ep$, $eA$) and $\mu$-LHC
($\mu p$, $\mu A$),
\item After the LHC (>2030): FCC ($pp$, $AA$), CLIC/LSC ($e^{-}$$e^{+}$),
PWFA-LC ($e^{-}$$e^{+}$), high energy MC ($\mu^{-}$$\mu^{+}$),
and FCC based lepton-hadron colliders, namely, $e$-FCC ($ep$, $eA$)
and $\mu$-FCC ($\mu p$, $\mu A$).
\end{itemize}
FCC is future 100 TeV center-of-mass energy $pp$ collider proposed
at CERN and supported by European Union within the Horizon 2020 Framework
Programme for Research and Innovation. Main parameters of the FCC
$pp$ option \cite{key-4} are presented in Table II. It includes
also an electron-positron collider option at the same tunnel (TLEP),
as well as several $ep$ collider options. Construction of the FCC
based $ep$ and $\mu p$ colliders will give opportunity to utilize
high(est) energy of proton beam for lepton-hadron collisions.

\begin{table}[H]
\caption{Main parameters of the FCC pp option.}

\centering{}%
\begin{tabular}{|c|c|}
\hline 
Beam Energy (TeV) & 50\tabularnewline
\hline 
\hline 
Peak Luminosity (10$^{34}$ cm$^{-2}$s$^{-1}$) & 5\tabularnewline
\hline 
Particle per Bunch ($10{}^{10}$) & 10\tabularnewline
\hline 
Transverse Emittance (rms, nm) & 2.2\tabularnewline
\hline 
$\beta^{*}$ amplitude function at IP (cm)  & 110-30\tabularnewline
\hline 
IP beam size ($\mu m$) & 6.8\tabularnewline
\hline 
Bunches per Beam  & 10600\tabularnewline
\hline 
Time between collisions ($\mu s$) & 0.025\tabularnewline
\hline 
Bunch Spacing (ns) & 25\tabularnewline
\hline 
Bunch Length (rms, mm) & 80\tabularnewline
\hline 
Beam-beam Tune Shift per crossing (10$^{-3}$) & 5-15\tabularnewline
\hline 
\end{tabular}
\end{table}

The scope of paper is following. In Section 2 we consider different
options for the FCC based $ep$ colliders and present luminosity estimations
for them. Main parameters of the FCC based $\mu p$ colliders are
considered in Section 3. Finally, Section 4 contains summary of obtained
results and recommendations.

\begin{figure}[H]

\begin{centering}
\includegraphics[scale=0.75]{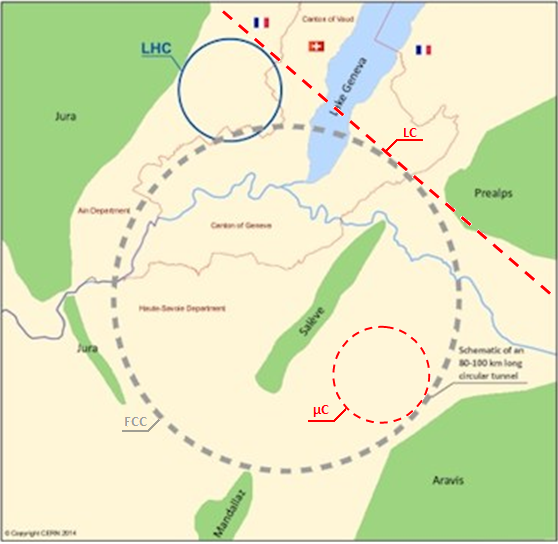}\caption{Possible configuration for FCC, linear collider (LC) and muon collider
($\mu C$).}

\par\end{centering}

\end{figure}

\section{FCC based $ep$ colliders}

As mentioned above FCC itself includes also $e^{-}$$e^{+}$ (TLEP)
and $ep$ collider options with $E_{e}=80$$,\:120$ and $175\:GeV$
(W-pair threshold, Higgs and t-pair threshold, respectively). In addition
the use of $E_{e}=60\:GeV$ conventional energy recovery linac (ERL60),
adopted as basic option for LHeC \cite{key-5}, for the FCC based
$ep$ collider is under consideration. One pass linac options (OPL)
for the FCC based $ep$ collider (see Fig 1), including versions with
second (decelerating) linac shoulder for energy recovery (OPERL),
have been considered in \cite{key-6}. In Table III we present main
parameters for the FCC based $ep$ colliders. Here we add also hypothetical
Linear Super Collider (see \cite{key-7,key-8} and references therein)
and as an extreme e-FCC case. For linac-ring type $ep$ colliders,
keeping in mind that at collision point e-beam transverse size is
smaller than p-beam transverse size, expression for luminosity can
be written as \cite{key-9}:

\begin{center}
$L_{ep}=\frac{1}{4\pi}\frac{P_{e}}{E_{e}}\frac{n_{p}}{\varepsilon_{p}^{N}}\frac{\gamma_{p}}{\beta_{p}^{\star}}\qquad(1)$ 
\par\end{center}

for round, transversely matched beams. Here $E_{e}$ and $P_{e}$
denote energy and beam power of electrons, respectively; $\gamma_{p}=E_{p}/m_{p}=5.33\times10^{4}$
(for rest of symbols see Table 2).

\begin{table}[H]
\caption{Main parameters of the FCC based $ep$ colliders}

\centering{}%
\begin{tabular}{|c|c|c|c|c|}
\hline 
Collider name & $E_{e}$ , TeV  & \textsurd s, TeV & $L_{ep}=10^{31}cm^{-2}s^{-1}$  & $L_{int}$, $fb^{-1}$ (per year)\tabularnewline
\hline 
\hline 
ERL60-FCC & 0.06 & 3.46 & 1000 \cite{key-10} & 100\tabularnewline
\hline 
FCC-e80 & 0.08 & 4.00 & 2300 \cite{key-10} & 230\tabularnewline
\hline 
FCC-e120 & 0.12 & 4.90 & 1200 \cite{key-10} & 120\tabularnewline
\hline 
FCC-e175 & 0.175 & 5.92 & 400 & 40\tabularnewline
\hline 
OPL500-FCC & 0.5 & 10.0 & 8 & 0.8 \textrightarrow{} 80\tabularnewline
\hline 
OPERL500-FCC & 0.5 & 10.0 & 20000 & 2000 \textrightarrow{} 200\tabularnewline
\hline 
OPL1000-FCC & 1 & 14.1 & 4 \cite{key-6} & 0.4 \textrightarrow{} 40\tabularnewline
\hline 
OPERL1000-FCC & 1 & 14.1 & 10000 \cite{key-6} & 1000 \textrightarrow{} 100\tabularnewline
\hline 
OPL5000-FCC & 5 & 31.6 & 0.8 & 0.08 \textrightarrow{} 8\tabularnewline
\hline 
OPERL5000-FCC & 5 & 31.6 & 2000 & 200 \textrightarrow{} 20\tabularnewline
\hline 
\end{tabular}
\end{table}

Luminosity values given in Table III assume simultaneous operation
with $pp$ collider. For OPL options these values can be increased
using dedicated proton beams with larger bunch population \cite{key-8}
(this opportunity is not efficient for e-ring and ERL options due
to beam-beam tune shift and disruption limitations, respectively).

The lower limit on $\beta_{p}^{\star}$, which is given by proton
bunch length, can be overcome by applying a \textquotedblleft dynamic\textquotedblright{}
focusing scheme \cite{key-11}, where the proton bunch waist travels
with electron bunch during collision. In this scheme $\beta_{p}^{\star}$
is limited, in principle, by the electron bunch length, which is two
orders magnitude smaller. More conservatively, an upgrade of the luminosity
by a factor of 3-4 may be possible.

An additional order of magnitude can be handled using cooling system
counteracting IBS of proton bunches \cite{key-12}. Combination of
three methods may give opportunity to handle two orders higher luminosity
values for all OPL options given in Table 3 (see last column).

Let us finish this section by following remark on OPERL version (Section
7.1.5 in \cite{key-5}): three orders of luminosity gain assumes overoptimistic
99.9\% energy recovery. For this reason we decrease values given in
last column of Table 3 by an order of magnitude.

\section{FCC based $\mu p$ colliders}

Muon-proton colliders were proposed almost 2 decades ago. Construction
of additional proton ring in \textsurd s = 4 TeV muon collider tunnel
was suggested in \cite{key-13} in order to handle $\mu p$ collider
with the same center-of-mass energy. However, luminosity value, namely
$L_{\mu p}=3\times10^{35}cm$$^{-2}s^{-1}$, was extremely over estimated,
realistic value for this option is three orders smaller \cite{key-8}.
Then, construction of additional 200 GeV energy muon ring in the Tevatron
ring in order to handle \textsurd s = 0.9 TeV $\mu p$ collider with
$L_{\mu p}=10^{32}cm$$^{-2}s^{-1}$ was considered in \cite{key-14}. 

In this paper we consider another design, namely, construction of
muon ring close to FCC (see Fig 1). For numerical calculation a basic
expression for the luminosity \cite{key-15}

\begin{center}
$L=f_{coll}\frac{n_{1}n_{2}}{4\pi\sigma_{x}\sigma_{y}}\qquad(2)$ 
\par\end{center}

has been used. For round beams this equation transfroms to

\begin{center}
$L_{pp}=f_{pp}\frac{n{}_{p}^{2}}{4\pi\sigma_{p}^{2}}\qquad(3)$ 
\par\end{center}

\begin{center}
$L_{\mu\mu}=f_{\mu\mu}\frac{n{}_{\mu}^{2}}{4\pi\sigma_{\mu}^{2}}\qquad(4)$ 
\par\end{center}

for FCC and MC, respectively. Concerning muon-proton collisions one
should use larger transverse beam sizes and smaller collision frequency
values. Keeping in mind that $f_{\mu\mu}$ is an orders smaller than
$f_{pp}$, following correlation between $\mu p$ and $\mu\mu$ luminosities
take place:

\begin{center}
$L_{\mu p}=(\frac{n_{p}}{n_{\mu}})(\frac{\sigma_{\mu}}{max[\sigma_{p},\,\sigma_{\mu}]})^{2}L_{\mu\mu}\qquad(5)$ 
\par\end{center}

Using parameters of $\mu\mu$ colliders given in Table IV \cite{key-16},
according to Eq. (5) we obtain parameters of the FCC based $\mu p$
colliders presented in Table V.

\begin{table}[H]
\caption{Muon collider parameters \cite{key-16}}

\begin{centering}
\begin{tabular}{|c|c|c|c|c|c|}
\hline 
\textsurd s, TeV & 0.126 & 0.35 & 1.5 & 3.0 & 6.0\tabularnewline
\hline 
Avg. Luminosity, $10^{34}cm^{-2}s^{-1}$ & 0.008 & 0.6 & 1.25 & 4.4 & 12\tabularnewline
\hline 
Circumference, km & 0.3 & 0.7 & 2.5 & 4.5 & 6\tabularnewline
\hline 
Repetition Rate, Hz & 15 & 15 & 15 & 12 & 6\tabularnewline
\hline 
$\beta^{\star}$, cm & 1.7 & 0.5 & 1  & 0.5  & 2.5\tabularnewline
\hline 
No. muons/bunch, $10^{12}$ & 4 & 3 & 2 & 2 & 2\tabularnewline
\hline 
No. bunches/beam & 1 & 1 & 1 & 1 & 1\tabularnewline
\hline 
Norm. Trans. Emmitance, $\pi\:mm-rad$ & 0.2 & 0.05 & 0.025 & 0.025 & 0.025\tabularnewline
\hline 
\end{tabular}
\par\end{centering}

\end{table}

\begin{table}[H]

\caption{Main parameters of the FCC based $\mu p$ colliders}

\centering{}%
\begin{tabular}{|c|c|c|c|c|}
\hline 
Collider name & $E_{\mu},\:TeV$ & \textsurd s, TeV & $L_{\mu p}=10^{31}cm^{-2}s^{-1}$ & $L_{int},\:fb^{-1}$(per year)\tabularnewline
\hline 
\hline 
$\mu63$-FCC & 0.063 & 3.50 & 0.2 & 0.02\tabularnewline
\hline 
$\mu175$-FCC & 0.175 & 5.92 & 20 & 2\tabularnewline
\hline 
$\mu750$-FCC & 0.75 & 12.2 & 50 & 5\tabularnewline
\hline 
$\mu1500$-FCC & 1.5 & 17.3 & 50 & 5\tabularnewline
\hline 
$\mu3000$-FCC & 3 & 24.5 & 300 & 30\tabularnewline
\hline 
\end{tabular}
\end{table}

Luminosity values presented in Table V assume simultaneous operation
with $pp$ collider. These values can be increased by an order using
dedicated proton beams with larger bunch population \cite{key-8}.

\section{$ep$ colliders based on the FCC and PWFA-LC}

Recently, muti-TeV CM energy $e^{-}$$e^{+}$ colliders based on Plasma
Wake-Field Acceleration Linear Collider (PWFA-LC) have been proposed
\cite{key-17}. In this section we estimate parameters of ep collisions
based on the FCC proton beam and PWFA-LC electron beam. For numerical
calculations we use parameters presented in Tables II and VI. 

The expression for luminosity of ep collisions is given by

\begin{flushright}
$L=\frac{N_{e}N_{p}}{4\pi\sigma_{p}^{2}}f_{c}\qquad\qquad\qquad\qquad\qquad\qquad\qquad\qquad\qquad\qquad(6)$ 
\par\end{flushright}

where $\sigma_{p}$ is IP proton beam size, N$_{e}$ and N$_{p}$
are electron and proton bunch population, f$_{c}$ is collision frequency.
We use $\sigma_{p}$ in Eq. (6) because electron beam size at IP is
much smaller. f$_{c}$ is determined by repetition rate of electron
beam. 

Beam-beam tune shift for proton beam is given by

\begin{flushright}
$\varDelta Q_{p}=\frac{N_{e}r_{p}\beta_{p}^{*}}{2\pi\gamma_{p}\sigma_{xe}(\sigma_{xe}+\sigma_{ye})}f_{c}\qquad\qquad\qquad\qquad\qquad\qquad\qquad\qquad\qquad\qquad(7)$ 
\par\end{flushright}

where r$_{p}$ is classical radius of proton, $\beta_{p}^{*}$ is
beta function of proton beam at interaction point, $\gamma_{p}$ is
the Lorentz factor of proton beam, $\sigma_{xe}$ and $\sigma_{ye}$
are horizontal and vertical beam sizes of electron, respectively.

Disruption parameter for electron beam is given by

\begin{flushright}
$D=\frac{2N_{p}r_{e}\sigma_{zp}}{\gamma_{e}\sigma_{xp}(\sigma_{xp}+\sigma_{yp})}f_{c}\qquad\qquad\qquad\qquad\qquad\qquad\qquad\qquad\qquad\qquad(8)$ 
\par\end{flushright}

where r$_{e}$ is classical radius of electron, $\gamma_{e}$ is the
Lorentz factor of electron beam, $\sigma_{xp}$, $\sigma_{yp}$ and
$\sigma_{zp}$ are horizontal and vertical beam sizes of proton and
bunch length of proton beam, respectively. In numerical calculations
we used matched electron and proton beams, namely $\sigma_{xe}$=$\sigma_{ye}$=$\sigma_{p}$.

Main parameters of the PWFA-LC and FCC based ep colliders are given
in Table VII, where upgraded FCC means N$_{p}$= 2.2 x 10$^{11}$,
$\beta_{p}^{*}=0.1$ m and $\sigma_{p}=2.05$ $\mu m$. In last two
columns we present N$_{p}$ and L$_{ep}$ values for limiting case
D$_{e}$ = 25.

\begin{table}[H]
\caption{PWFA-LC electron beam parameters.}

\centering{}%
\begin{tabular}{|c|c|c|c|c|c|}
\hline 
Beam Energy (GeV) & 125 & 250 & 500 & 1500 & 5000\tabularnewline
\hline 
\hline 
Peak Luminosity (10$^{34}$ cm$^{-2}$s$^{-1}$) & 0.94 & 1.25 & 1.88 & 3.76 & 6.27\tabularnewline
\hline 
Particle per Bunch ($10{}^{10}$) & 1 & 1 & 1 & 1 & 1\tabularnewline
\hline 
Norm. Horizontal Emittance (m) & 1.00$\times10^{-5}$ & 1.00$\times10^{-5}$ & 1.00$\times10^{-5}$ & 1.00$\times10^{-5}$ & 1.00$\times10^{-5}$\tabularnewline
\hline 
Norm. Vertical Emittance (m) & 3.50$\times10^{-8}$ & 3.50$\times10^{-8}$ & 3.50$\times10^{-8}$ & 3.50$\times10^{-8}$ & 3.50$\times10^{-8}$\tabularnewline
\hline 
Horizontal beam size at IP ($m$) & 6.71$\times10^{-7}$ & 4.74$\times10^{-7}$ & 3.36$\times10^{-7}$ & 1.94$\times10^{-7}$ & 1.06$\times10^{-7}$\tabularnewline
\hline 
Vertical beam size at IP ($m$) & 3.78$\times10^{-9}$ & 2.67$\times10^{-9}$ & 1.89$\times10^{-9}$ & 1.09$\times10^{-9}$ & 5.98$\times10^{-10}$\tabularnewline
\hline 
Bunches per Beam  & 1 & 1 & 1 & 1 & 1\tabularnewline
\hline 
Repetition Rate ($Hz$) & 30000 & 20000 & 15000 & 10000 & 5000\tabularnewline
\hline 
Beam Power at IP (MW) & 6 & 8 & 12 & 24 & 40\tabularnewline
\hline 
Bunch Spacing (ns) & 3.33$\times10^{4}$ & 5.00$\times10^{4}$ & 6.67$\times10^{4}$ & 1.00$\times10^{5}$ & 2.00$\times10^{5}$\tabularnewline
\hline 
Bunch Length at IP (m) & 2.00$\times10^{-5}$ & 2.00$\times10^{-5}$ & 2.00$\times10^{-5}$ & 2.00$\times10^{-5}$ & 2.00$\times10^{-5}$\tabularnewline
\hline 
Disruption  & 8.44$\times10^{-1}$ & 2.39$\times10^{-1}$ & 6.71$\times10^{-1}$ & 3.51 & 21.4\tabularnewline
\hline 
\end{tabular}
\end{table}

\begin{table}[H]
\caption{PWFA-LC-FCC parameters.}

\begin{centering}
\begin{tabular}{|c|c|c|c|c|c|c|c|}
\hline 
 &  & \multicolumn{2}{c|}{Nominal FCC} & \multicolumn{2}{c|}{Upgraded FCC} & \multicolumn{2}{c|}{D$_{e}$ = 25}\tabularnewline
\hline 
\hline 
$E_{e}$ (GeV) & $\sqrt{s}$ (TeV) & L, 10$^{30}cm^{-2}s^{-1}$ & D & L, 10$^{30}cm^{-2}s^{-1}$ & D & N$_{p}$(10$^{11}$) & L, 10$^{30}cm^{-2}s^{-1}$\tabularnewline
\hline 
125 & 5.00 & 5.16 & 1.99 & 124 & 47.6 & 1.1 & 62\tabularnewline
\hline 
250 & 7.08 & 3.44 & 1.00 & 82.6 & 24 & 2.2 & 82.6\tabularnewline
\hline 
500 & 10.0 & 2.58 & 0.50 & 61.9 & 12 & 4.4 & 124\tabularnewline
\hline 
1500 & 17.3 & 1.72 & 0.17 & 41.3 & 4.1 & 12 & 240\tabularnewline
\hline 
5000 & 31.6 & 0.86 & 0.05 & 20.8 & 1.2 & 44 & 400\tabularnewline
\hline 
\end{tabular}
\par\end{centering}

\end{table}

One can see from Table III that luminosity of ep collisions of order
of 10$^{32}$ cm$^{-2}$ s$^{-1}$ is achievable for all PWFA-LC stages.
In principle luminosity values may be increasing by an order using
dynamic focusing for proton beams \cite{key-11,key-12}.

\section{Conclusion}

The FCC based $ep$ and $\mu p$ colliders will provide opportunity
to achieve multi-TeV center-of-mass energy scale at partonic level
in lepton-hadron collisions with sufficiently high luminosities. Summary
of main parameters of these machine, which can be used by our colleagues
for research of physics search potential of e-FCC and $\mu$-FCC,
is given in Table VI. As an example, FCC based $ep$ ($\mu p$) colliders
have a great potential for the first (second) family leptoquarks and
color octet electron (muon) search. There are a lot of BSM phenomena
which can be investigated in the best manner at multi-TeV scale lepton-hadron
colliders. Finger estimations show that BSM physics search potential
of ep colliders is comparable to that of FCC and essentially exceeds
that of corresponding e$^{+}$e$^{-}$ collider (for comparison of
the LHC pp, ILC e$^{+}$e$^{-}$ and ILC-LHC ep search potentials
see \cite{key-3} and references therein). Moreover, these machines
will provide opportunity to investigate extremely small x-Bjorken
region, which is crucial for clarifying the QCD basics, as well as
the origin of 98\% of mass of visible universe \cite{key-18}.

\begin{table}[H]
\caption{Summary of main parameters of the FCC based $lp$ colliders}

\centering{}%
\begin{tabular}{|c|c|c|c|}
\hline 
Collider name & $E_{l},\:TeV$ & \textsurd s, TeV & $L_{int},\:fb^{-1}$(per year)\tabularnewline
\hline 
\hline 
ERL60-FCC & 0.06 & 3.46 & 100\tabularnewline
\hline 
FCC-e80 & 0.08 & 4.00 & 230\tabularnewline
\hline 
FCC-e120 & 0.12 & 4.90 & 120\tabularnewline
\hline 
FCC-e175 & 0.175 & 5.92 & 40\tabularnewline
\hline 
OPL500-FCC & 0.5 & 10.0 & 10-100\tabularnewline
\hline 
OPERL500-FCC & 0.5 & 10.0 & 100-300\tabularnewline
\hline 
OPL1000-FCC & 1 & 14.1 & 5-50\tabularnewline
\hline 
OPERL1000-FCC & 1 & 14.1 & 50-150\tabularnewline
\hline 
OPL5000-FCC & 5 & 31.6 & 1-10\tabularnewline
\hline 
OPERL5000-FCC & 5 & 31.6 & 10-30\tabularnewline
\hline 
$\mu63$-FCC & 0.063 & 3.50 & 0.1-1\tabularnewline
\hline 
$\mu175$-FCC & 0.175 & 5.92 & 2-20\tabularnewline
\hline 
$\mu750$-FCC & 0.75 & 12.2 & 5-50\tabularnewline
\hline 
$\mu1500$-FCC & 1.5 & 17.3 & 5-50\tabularnewline
\hline 
$\mu3000$-FCC & 3 & 24.5 & 10-100\tabularnewline
\hline 
PWFA125-FCC & 0.125 & 5 & 1-10\tabularnewline
\hline 
PWFA250-FCC & 0.25 & 7.08 & 1-10\tabularnewline
\hline 
PWFA500-FCC & 0.5 & 10.0 & 1-10\tabularnewline
\hline 
PWFA1500-FCC & 1.5 & 17.3 & 2-20\tabularnewline
\hline 
PWFA5000-FCC & 5 & 31.6 & 4-40\tabularnewline
\hline 
\end{tabular}
\end{table}

It should be noted that OPL/OPERL-FCC and PWFA-LC-FCC $ep$ colliders
will give opportunity to construct also $\gamma p$ colliders with
approximately same center-of-mass energy and luminosity. In addition
approval of the FCC $AA$ collider option will give opportunity to
handle also multi-TeV energy $\gamma A$ and $\mu A$ collisions (see
review \cite{key-8} and references therein). Also FEL $\gamma A$
option has a great potential for nuclear spectroscopy. These options
are under consideration \cite{key-19}.
\begin{acknowledgments}
Authors are gratefull to Frank Zimmermann for useful discussions.
This work is supported by TUBITAK under the grant no 114F337.\end{acknowledgments}

\end{document}